\documentstyle[11pt,newpasp,twoside,psfig]{article}
\begin{document}

\title{Ionized Disc Models of MCG--6-30-15}
\author{D.R. Ballantyne, A.C. Fabian and S. Vaughan}
\affil{Institute of Astronomy, Madingley Road, Cambridge,
U.K. CB3~OHA}

\index{Ballantyne, D.R.}
\index{Fabian, A.C.}
\index{Vaughan, S.}

\begin{abstract}
Ionized reflection is one of the possible mechanisms proposed to
explain the small short-timescale variability of the MCG--6-30-15
Fe~K$\alpha$ line. In this contribution, we fit the Ross \& Fabian
ionized reflection models to the 325~ks {\it XMM-Newton} observation
of MCG--6-30-15. It is found that the spectrum above 2.5~keV can be
fitted with a combination of an ionized reflector from 5~$r_g$ and a
neutral reprocessor beyond $\sim$70~$r_g$. The ionization parameter of the
inner reflector increases as the source brightens.  
\end{abstract}

\section{Introduction}
MCG--6-30-15 has a strong and very broad Fe~K$\alpha$ line which shows
little short timescale variability (Fabian et al. 2002). A possible
explanation for this behavior is if the line and reflection continuum
arises from an ionized accretion disc. At ionization parameters $\log
\xi > 3$, where $\xi=4\pi F_{\mathrm{X}}/n_{\mathrm{H}}$, the Fe K
line is produced by He-like iron, and its equivalent width decreases
as $\xi$ increases (Ross, Fabian \& Young 1999). In this contribution
we present the results of fitting ionized
disc models to a 325 ks observation of MCG--6-30-15 made
simultaneously by {\it XMM-Newton} and {\it BeppoSAX}. A complete
description of the analysis can be found in the paper by Ballantyne,
Vaughan \& Fabian (2002).

\section{Data and Models}
The data has been described by Fabian et al. (2002). Here, we present
results from fitting the EPIC pn, MECS and PDS data between 2.5 and
80~keV. The low energy cutoff was chosen to avoid spectral complexity due to
the warm absorber, while the 100~keV upper-limit of the reflection
models set the value of the high energy cutoff. The constant density
ionized disc models of Ross \& Fabian (1993) were used in the spectral
fitting (done with \textsc{xspec}~v.11.2). Relativistic blurring was
taken into account, as was Galactic absorption.  

\section{Results}
The Fe line exhibited by MCG--6-30-15 is very strong (EW $\approx
450$~eV) and broad (red wing extends to $\sim 3$~keV). It was
difficult for a single ionized reflector to account for both of these
features of the Fe line and fit the continuum, even with an
overabundance of iron. Therefore, we were forced to consider 2
reflectors to fit the line (see left panel of Fig.~1).
\begin{figure}[t]
\centerline{
\psfig{figure=ballantyne_fig1.ps,width=0.5\textwidth,angle=-90}
\psfig{figure=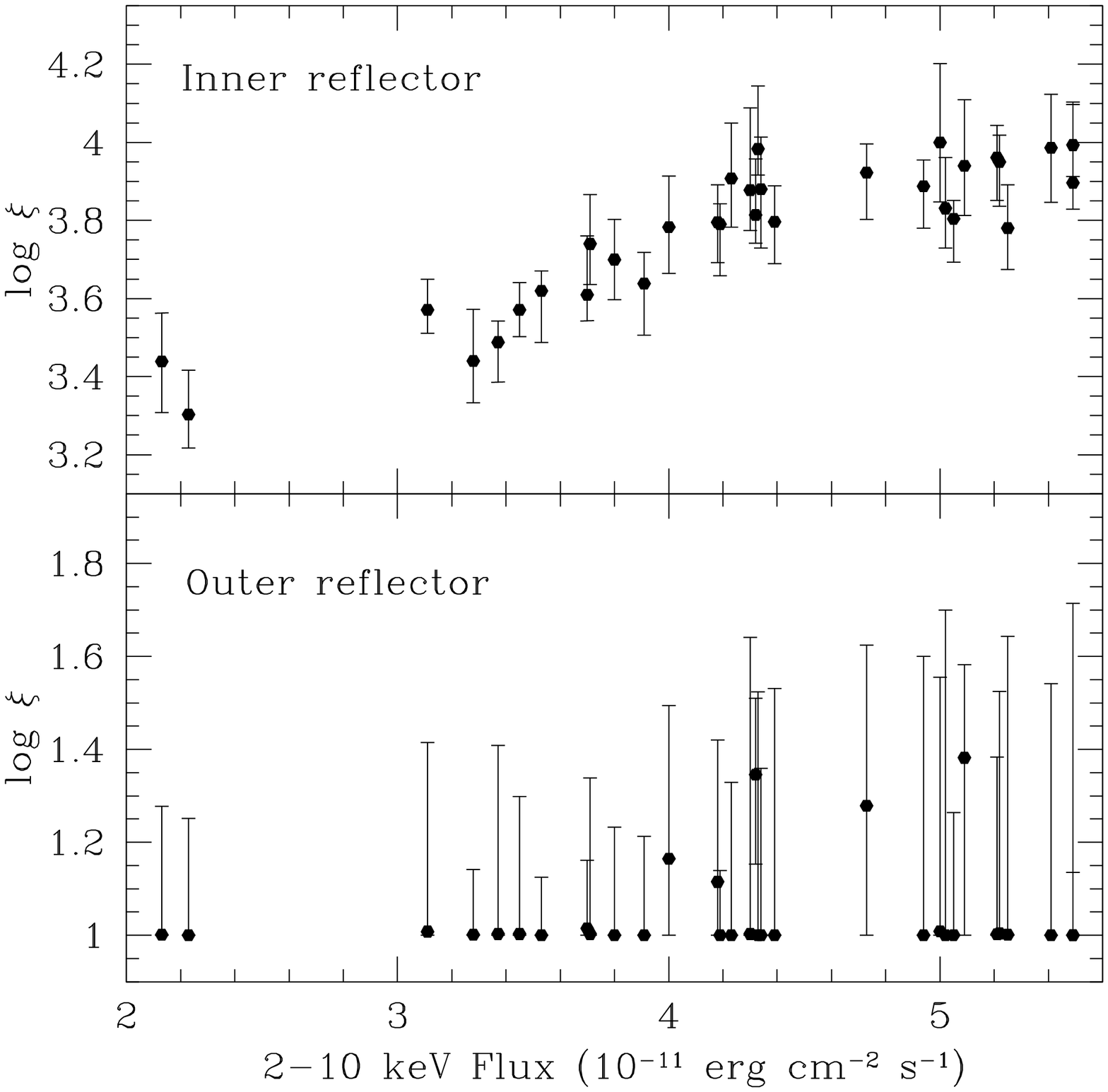,width=0.5\textwidth}
}
\caption{(Left) The time-averaged double reflection model that best
fits the data between 2.5 and 80~keV. Galactic absorption has been
removed. The solid line is the full model; the dashed line denotes
the inner, ionized reflector ($\log \xi=3.82$, $r_{in}=4.9$~$r_g$,
$r_{out}=5.0$~$r_g$); the incident $\Gamma=1.92$ power-law is shown
with the dotted line, and the outer, neutral reprocessor ($\log
\xi=1.0$, $r_{in}=71$~$r_g$, $r_{out}=1822$~$r_g$) is the dot-dashed
line. (Right) Evolution of $\xi$ for the two components with the total
2--10~keV flux.}
\label{fig1}
\end{figure}
In this model, the red wing of the Fe line and the bulk of the
2--10~keV continuum were fit by an ionized ($\log \xi=3.8$) reflection
component and the primary power-law (with the reflection
fraction frozen at 1). Oddly, this component is constrained to arise
from a narrow annulus at $\sim$5~$r_g$. The blue horn of the Fe line
and the majority of the high energy continuum is produced by
secondary reflection of the primary power-law beyond $\sim$70~$r_g$. 

The right panel of Fig.~1 shows how the ionization parameters of the
two reflectors changed with the source brightness. These fits were
made with 10~ks segments from all three orbits of the {\it XMM-Newton}
pn dataset. We found that the inner reflector becomes more ionized as
MCG--6-30-15 brightened, while the secondary remained
neutral. Therefore, the red wing of the Fe line weakens in higher flux
states, but the blue core will respond on much longer timescales. In
this way, a double reflection model can account for some of the Fe
line variability properties.



\begin{references}
\reference Ballantyne, D.R., Vaughan, S. \& Fabian, A.C., 2002,
\mnras, submitted
\reference Fabian, A.C., et al., 2002, \mnras, 335, L1
\reference Ross, R.R. \& Fabian, A.C., 1993, \mnras, 261, 74
\reference Ross, R.R., Fabian, A.C. \& Young, A.J., 1999, \mnras, 306, 461

\end{references}
\end{document}